\documentclass[11pt]{article}
\usepackage{moriond,epsfig}

\def\Journal#1#2#3#4{{#1} {\bf #2}, #3 (#4)}

\def\NIMA{{\em Nucl. Instrum. Methods} A}

\def\PLB{{\em Phys. Lett.}  B}
\def\PRL{\em Phys. Rev. Lett.}
\def\PRD{{\em Phys. Rev.} D}

\def\ra{\rightarrow}

\def\be{\begin{equation}}
\def\ee{\end{equation}}
\def\bea{\begin{eqnarray}}
\def\eea{\end{eqnarray}}

\newcommand{\pio}{\mbox{$\pi^{0}$}}
\newcommand{\pim}{\mbox{$\pi^{-}$}}
\newcommand{\pip}{\mbox{$\pi^{+}$}}

\newcommand{\ks}{\mbox{$K_{S}$}}
\newcommand{\kl}{\mbox{$K_{L}$}}
\newcommand{\kp}{\mbox{$K^{+}$}}
\newcommand{\km}{\mbox{$K^{-}$}}

\newcommand{\br}{\mbox{${\rm BR}$}}

\begin{document}
\vspace*{4cm}
\title{HIGHLIGHTS OF THE KLOE EXPERIMENT AT DA$\Phi$NE}

\author{ The KLOE Collaboration 
\footnote{The KLOE Collaboration:
A.~Aloisio,
F.~Ambrosino,
A.~Antonelli, 
M.~Antonelli,
C.~Bacci,
G.~Bencivenni,
S.~Bertolucci,
C.~Bini,
C.~Bloise,
V.~Bocci
% ...
} 
\\
\vspace{1.0cm} presented by STEFANO MISCETTI }

\address{ INFN Laboratori Nazionali di Frascati, Via Enrico Fermi 40, 
00044 RM, Italy}

\maketitle\abstracts{
The KLOE experiment at DA$\Phi$NE has collected $\sim$ 450 pb$^{-1}$
of $e^{+}e^{-}$ collisions at center of mass energy $W \sim 1.02$ GeV. 
Preliminary results are presented for the most recent measurements: 
limit on the BR($\ks \ra 3 \pio$), BR of the
$K_{e3}$ decay of the $\ks$ and determination of the hadronic
cross section.}

\section{Introduction}
DA$\Phi$NE, the Frascati $\phi$ factory, is an $e^{+}e^{-}$ collider
working at $W \sim m_{\phi} \sim 1.02$ GeV with a design luminosity
of $5 \cdot 10^{32}$ cm$^{-2}$ s$^{-1}$. $\phi$ mesons are produced,
essentially at rest, with a visible cross section of $\sim$ 3.2 $\mu$b
and decay into $\kp\km$ ($\ks\kl$) pairs with BR of $\sim 49$\% 
($\sim 34$\%).
These pairs are produced in a pure $J^{PC}=1^{--}$ quantum state, so that 
observation of a $\ks$ ($\kp$)  in an event signals (tags) the presence of a $\kl$
($\km$) and viceversa; highly pure and nearly monochromatic $\ks,\kl,\kp$ and $\km$
beams can be obtained. Neutral kaons get a momentum of $\sim$ 110 MeV/c which
translates in a slow speed, $\beta_{K} \sim$ 0.22.
$\ks$ and $\kl$ can therefore be distinguished by their mean decay lengths:
$\lambda_{S} \sim $ 0.6 cm and $\lambda_{L} \sim $ 340 cm.
 
The KLOE detector consists essentially of a  drift chamber, DCH, surrounded by an
electromagnetic calorimeter, EMC. The DCH~\cite{nimdch} is a cylinder of 4 m diameter
and 3.3 m in length which constitutes a large fiducial volume 
for $\kl$ decays (1/2 of $\lambda_{L}$). The momentum resolution for tracks 
at large polar angle is $\sigma_{p}/p \leq 0.4$\%. The EMC is a 
lead-scintillating fiber calorimeter~\cite{nimcalo}
consisting of a barrel and two endcaps which cover 98\% of the solid angle. The
energy resolution is $\sigma_{E}/E \sim 5.7\%/\sqrt{\rm{E(GeV)}}$. The intrinsic
time resolution is $\sigma_{T} =$ 54 ps$/\sqrt{\rm{E(GeV)}} \oplus 50$ ps. 
A superconducting coil surrounding the barrel provides a 0.52 T magnetic field.

During 2002 data taking, the maximum luminosity reached by DA$\Phi$NE was 
$7.5 \cdot 10^{31}$ cm$^{-2}$ s$^{-1}$. Although this is lower than the design value,
the performance of the machine was improving during the years and, at the end
of 2002, we collected $\sim$ 4.5 pb$^{-1}$/day. The whole sample (2001-2002)
amounts to 450 pb$^{-1}$, equivalent to 1.4 billion $\phi$ decays. Recently, 
the machine has been upgraded and KLOE is resuming its data taking in 
spring 2004.

\section{Kaon physics}

The tagging of $\kl$ and $\ks$ is the basis of each KLOE analysis 
for neutral kaons. Similar techniques have been developed also 
for charged kaons. The selection of $\ks \ra \pip \pim$ decays 
provides an efficient tag for $\kl$ decays. $\ks$'s  are instead 
tagged by identifying  a $\kl$ interaction, $\kl$-crash, 
in the calorimeter,
which has a very distinctive signature given by a late 
($\beta_K = 0.2)$ high-energy cluster not 
associated to any track. In either case, reconstruction of one kaon 
establishes the trajectory of the other one with an angular resolution 
of 1$^{\circ}$ and a momentum resolution of $\sim$ 2 MeV. 
Several analyses~\cite{ks2pi,klgg} have been already completed or are under 
completion at KLOE. We discuss only the two most advanced items in progress.

\subsection{ Direct search of $\ks \ra 3 \pio$ }

The decay $\ks \ra 3 \pio$  is a pure CP violating process. 
The related CP violation parameter $\eta_{000}$ is defined as the 
ratio of \ks\ to \kl\ decay amplitudes:
$\eta_{000} = A(\ks \ra 3 \pio)/ A( \kl \ra 3 \pio)=
\varepsilon + \varepsilon^{'}_{000}$
where $\varepsilon$ describes the CP violation in the mixing matrix 
and $\varepsilon^{'}_{000}$ is a direct CP violating term. 
In the Standard Model we expect  $\eta_{000}$ to 
be similar to  $\eta_{00}$. 
The expected branching ratio of this decay
is therefore $\sim 2 \cdot 10^{-9}$, making its direct 
observation really challenging.
The best upper limit on the BR (i.e. on $|\eta_{000}|^2$) has been set 
to $1.5 \cdot 10^{-5}$ by SND~\cite{SND3pi0} where, similar to KLOE, 
it is possible to tag a $\ks$ beam. The other existing technique
is to detect the interference term between $\ks \kl $ in the same
final state which is proportional to $\eta_{000}$. The weighted
average of the best published values~\cite{BARMIN,CPLEAR3pi0} 
gives: 
$\eta_{000} =  (0.08 \pm 0.11) + i\cdot (0.07 \pm 0.16)$.
Apart from the interest in observing this decay directly, 
the large uncertainty on $\eta_{000}$ limits the precision on CPT
invariance test via the unitarity relation~\cite{Zhou}. 
In the most general way, a neutral kaon state~\cite{buchanan} is
expressed as:
$K_{S,L} = K_{1,2} + (\varepsilon \pm \delta) K_{2,1}$
where $K_1$ and $K_2$ are the two CP eigenstates and $\delta$
is a CPT violation parameter.
%
The unitarity relation in this base can be written as:
\begin{equation}
(1+i\tan(\phi_{sw})) (\Re(\varepsilon) -i \Im(\delta)) 
= \sum ( A^{*}(\ks \ra f) A(\kl \ra f)/\Gamma_{S})
\label{eq:unitarity}
\end{equation}
\noindent where the sum runs over all the possible decay channels $f$,
and $\tan(\phi_{sw})= 2\Delta m_{S,L}/(\Gamma_{S}-\Gamma_{L})$.
According to ref.~\cite{CPLEARcpt}, the value of 
$\Im(\delta)=(2.4 \pm 5.0)\cdot 10^{-5}$ 
is limited by  the measurement on $\eta_{000}$.  
Neglecting this term, 
the same analysis yields $\Im(\delta)=(-0.5 \pm 2.0)\cdot 10^{-5}$.

Our selection starts by requiring a $\kl$-crash tag and six
neutral clusters coming from the interaction 
point, IP. A tight constraint on $\beta$ and moderate requirements on
energy and angular acceptance are applied in order to
have a large control sample for the  background, while
retaining large selection efficiency for the signal.
On 450 pb$^{-1}$  we have an initial sample of 39\,k events 
dominated by $\ks \ra 2 \pio + $ 2 fake $\gamma$.
To  reduce the sample, a kinematic fit which imposes $\ks$ mass,
$\kl$ 4-momentum conservation and $\beta=1$ 
for each $\gamma$ is applied. Only the events with 
$\chi^{2}_{\rm{fit}}/\rm{ndf} < 3$ are retained for
further analysis.
However, this cut improves the rejection power only of a factor $\sim$ 3
and, to better discriminate $2 \pio$ {\it vs} $3 \pio$ final state, we
build two pseudo-$\chi^2$ variables:
$\chi^2_{3\pi}$, which is based on the 3 best $\pio$ mass 
estimates 
and $\chi^2_{2\pi}$, which selects 4 out of the 6 photons  providing
the best kinematic agreement with the considered decay.

\begin{figure}[!t]
\begin{center}
\begin{tabular}{cc}
\epsfig{figure=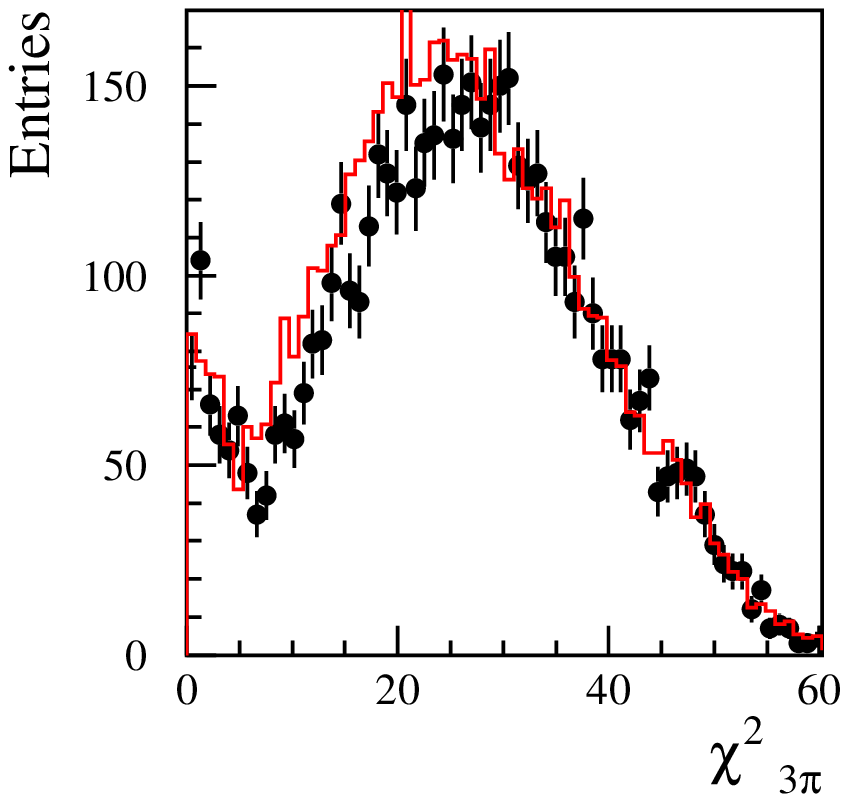,width=7.8cm}
\epsfig{figure=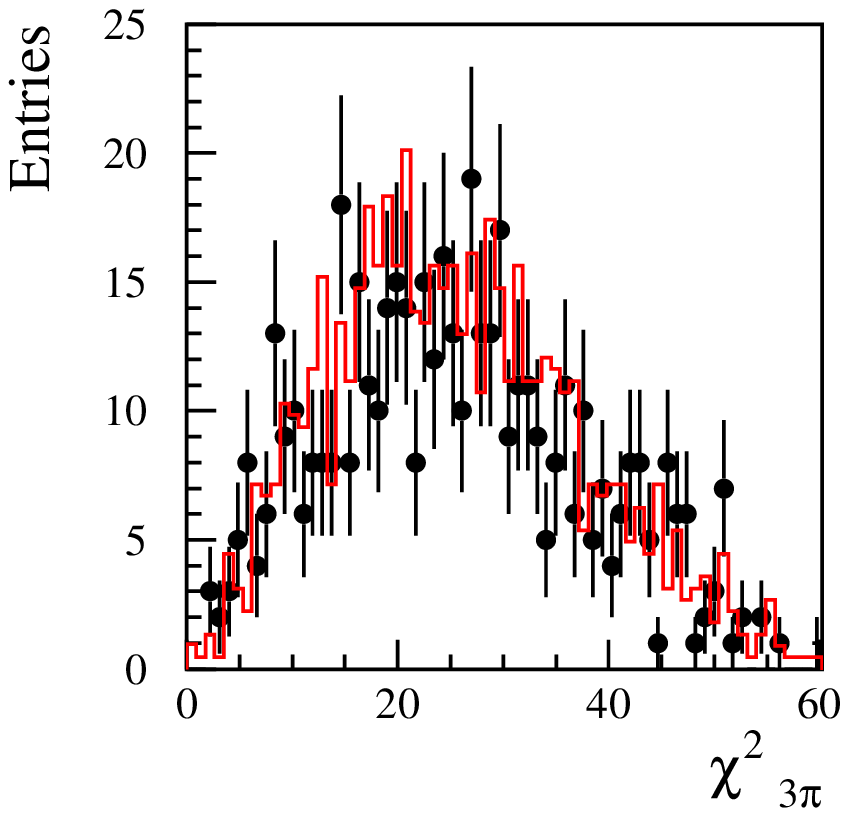,width=7.8cm}
\end{tabular}
\end{center}
\caption{Distribution of $\chi^2_{3\pi}$ when $14<\chi^2_{2\pi}<40$:
(left) total sample after acceptance selection, 
(right) all analysis cuts applied; black
dots (solid line) are data (MonteCarlo).
\label{fig:chi3pi}}
\end{figure}
 
The distribution of $\chi^2_{3\pi}$ is
shown in Fig.~\ref{fig:chi3pi}.a for the whole 
preselected sample by requiring $\chi^2_{2\pi}$ to be 
in a high acceptance region for the signal.
The presence of the large peak, at low $\chi^2_{3\pi}$ values,
indicates another source of contamination related to the
production of fake $\kl$-crash followed by a $\kl \ra 3 \pio$ decay. 
These fake, late clusters are produced by the pions from $\ks \ra \pip \pim$
interacting on the quadrupoles. Our MonteCarlo, MC, reproduces well
this background source (3\% of the total rate). To reduce it to a 
negligible amount we veto events with tracks coming  from the IP.
A signal box region in the $\chi^2_{2\pi}$ {\it vs} $\chi^2_{3\pi}$ plane
has been defined by optimizing  the upper limit in the MC sample.
With an efficiency $\varepsilon_{3\pi} = (22.6 \pm 0.8 $)\%,
we count 4 events for an expected background 
$N_b = 3 \pm 1.4 \pm 0.2 $. 
%
Folding the proper background uncertainty, we  quote the number of 
$\ks \ra 3 \pio$ decay
to be below 5.8 at 90\% C.L.
In the same tagged sample, we count $3.8\cdot 10^7$ $\ks \ra 2 \pio$ events
used for normalization. We finally derive
$\br(\ks \ra 3 \pio) \leq 2.1 \cdot 10^{-7}$ at 90\% C.L.\
which improves of a factor $\sim$ 100 the previous measurement.
%
This result can also be translated into a
limit  $|\eta_{000}|< 0.024$ at 90\% C.L. which makes 
the contribution of the uncertainty for this decay 
negligible in the calculation of $\Im(\delta)$.

\subsection{ Semileptonic decays and $V_{us}$}
The semileptonic charge asymmetries for $K_{S,L}$ are related to the CP, CPT 
violation parameters $\varepsilon,\,\delta$  as~\cite{buchanan,dambrosio} :
$A_{S,L} = \frac{\Gamma_s (\pi^+ e^- \overline{\nu} )-\Gamma_s(\pi^- e^+ \nu)}
{\Gamma_s (\pi^+ e^- \overline{\nu} )+\Gamma_s(\pi^- e^+ \nu)} = 
2 \Re (\varepsilon) \pm  2\Re(\delta)$.
A non zero value of $A_S-A_L$ would signal CPT violation
either in the $K_S$-$K_L$ mixing or in direct transitions 
violating the $\Delta S = \Delta Q$ rule. While $A_{L}$ is measured with high
precision~\cite{KTeV} a measurement of $A_S$ is still not existent.
KLOE has already measured~\cite{ksemilep} the BR for the $K_{e3}$ decay of the $\ks$
using 17 pb$^{-1}$ collected in 2000. A new measurement 
with the collected statistics of 450 pb$^{-1}$ gives a first 
determination of $A_S$. 
Moreover, a precise determination of $\Gamma_s(\pi e \nu)$  permits us
to evaluate $V_{us}$.

\begin{figure}[!t]
\begin{tabular}{cc}
\epsfig{figure=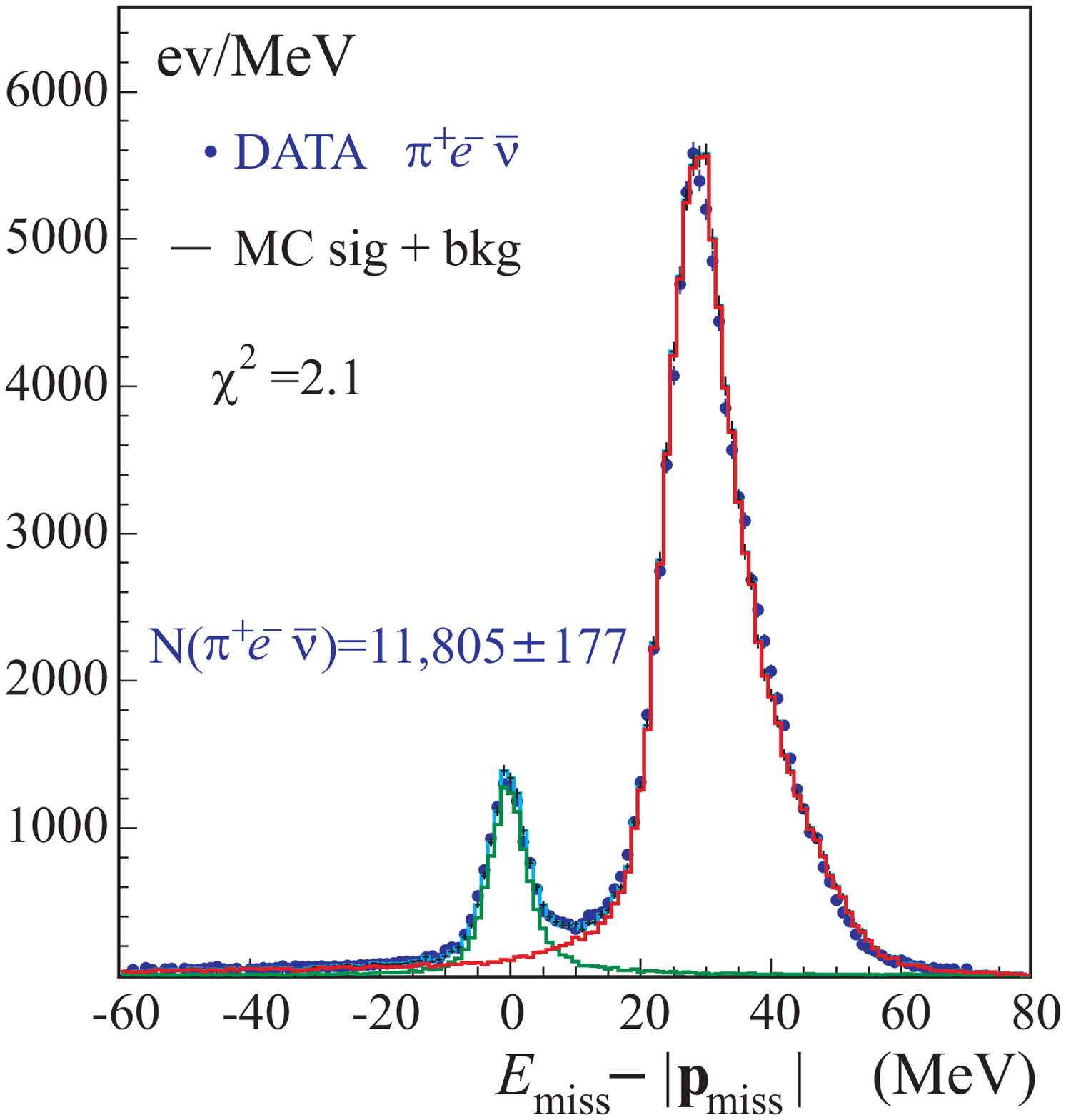,width=7.0cm}
\epsfig{figure=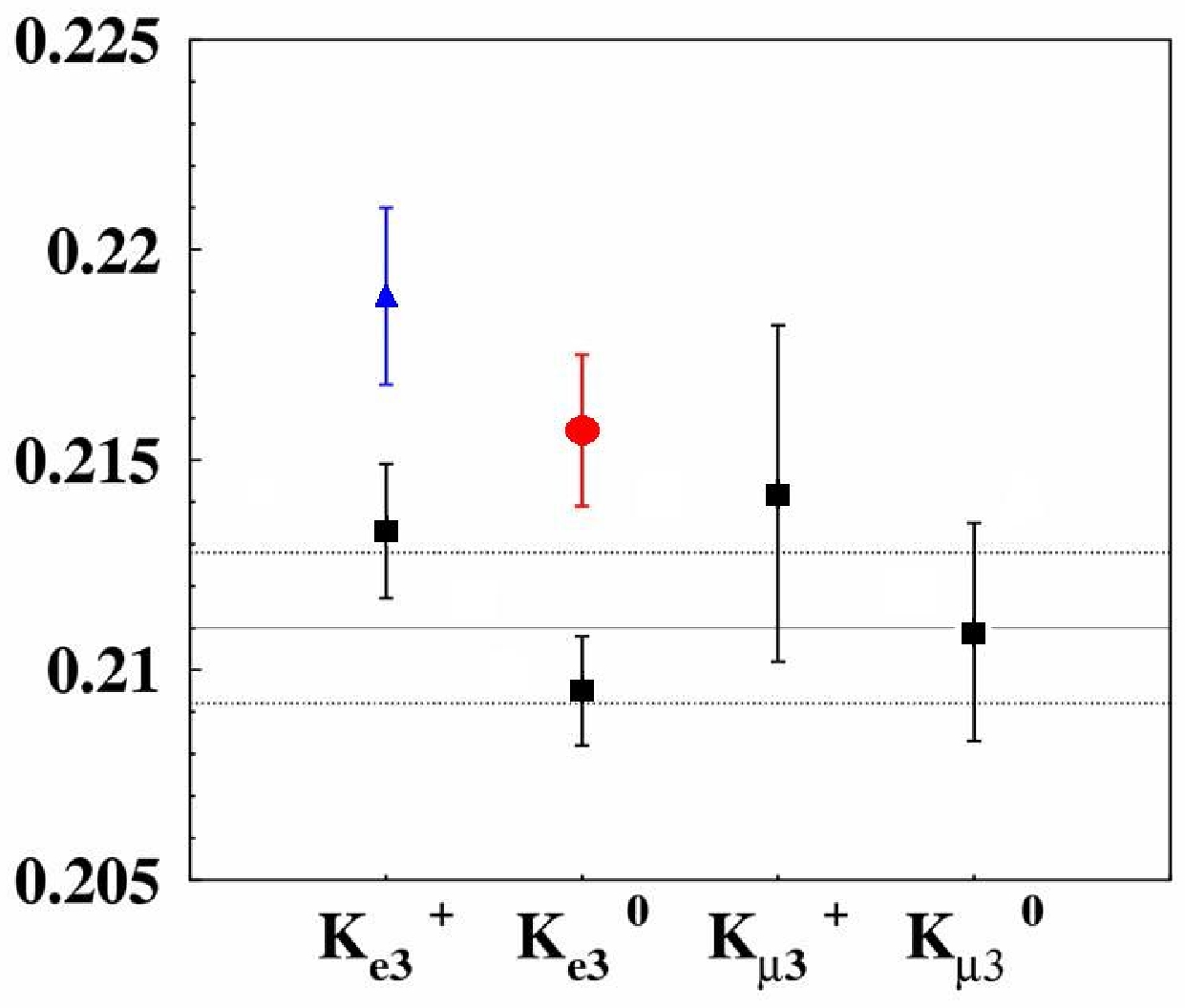,width=9.0cm}
\end{tabular}
\caption{(Left) $E_{\rm miss}-|P_{\rm miss}|$ distribution after time-of-flight
cuts for the $\pi^+ e^- \overline\nu$ sample: dots are data, solid lines
are MonteCarlo expectations. (Right) evaluations of 
$|V_{us}f_{+}^{K^0 \pi^-}(0)|$ from PDG 2002 numbers (squares), 
BNL-E865 (triangle), KLOE preliminar number from $K_{e3}$ of 
the $\ks$ (circle).}
\label{fig:semilep}
\end{figure}

The $K_S \rightarrow \pi e \nu$ decays are selected, after $\kl$-crash tagging, 
by the presence of two oppositely charged tracks from a vertex close to the IP. 
Loose momentum and angular cuts, and the requirement of an upper cut on 
$M(\pip\pim)$, reject most of the $\ks \ra \pip \pim$ background. 
The $\pi$ and $e$  assignments are made using time-of-flight 
so that the BR's to final states of each lepton charge can 
be measured independently. 
In Fig.~\ref{fig:semilep}.a, the $E_{\rm miss}-|P_{\rm miss}|$ distribution,
obtained by using the $\ks$ momentum estimated from the $\kl$-tag,
shows a pronounced peak around zero due to the neutrino.
The number of signal events is obtained from a fit 
which uses the MC distributions for signal and background with
their normalizations as free parameters. The generator used 
for the signal properly handles the final state emitted radiation through an 
infrared finite treatment.
By normalizing to the number of $\ks \ra \pip \pim$ events
counted in the same tagged sample, we get the following 
preliminary values  for 
BR$(\ks \ra \pi^+ e^- \overline{\nu})=(3.54 \pm 0.05 \pm 0.04)\cdot 10^{-4}$ and
BR$(\ks \ra \pi^- e^+ \nu ) = (3.54 \pm 0.05 \pm 0.04)\cdot 10^{-4}$.
Without considering the charge, we get 
BR$(\ks \ra \pi e \nu) = (7.09 \pm 0.07 \pm 0.08)\cdot 10^{-4}$,
which is consistent with our old measurement, improving of a
factor 5  the statistical error. 
On the basis of these results, we derive also the first 
measurement ever done of the charge asymmetry for the $\ks$:
$A_S= (-2 \pm 9 \pm 6)\cdot 10^{-3}$. This value is consistent with
the much more precise $A_{L}$ evaluations. With the 2 fb$^{-1}$
expected from next running we could perform a consistency test 
of $A_S$ with $2\Re(\varepsilon)$. We need instead at least 20 fb$^{-1}$
to determine $\delta$ with a precision comparable to the one obtained 
by CPLEAR.

The determinations of $|V_{us}|$ and $|V_{ud}|$ provide the most
precise test of CKM unitarity: $(|V_{ud}|^2 + |V_{us}|^2 = 1-\Delta$).
In PDG 2002~\cite{pdg02},  $\Delta= 0.0042 \pm 0.0019$
shows a $2.2\,\sigma$ deviation from unitarity.
In this test, $|V_{us}|$ account for 0.0011 of the error and
is derived from the measurement
of partial widths~\cite{CKMwsp} in $K_{l3}$ decays:
\begin{equation}
\Gamma(K_{l3}) \propto |V_{us} f_{+}^{K^0\pi^-}(0)|^2 I(\lambda_{+},\lambda_{0},0)
(1+\delta_{SU2} + \delta_{k})
\label{eq:semilep}
\end{equation}
\noindent where $f_{+}^{K^0\pi^-}(0)$ is the kaon form factor 
a $t=(P_k-P_\pi)^2=0$, 
$\lambda_{+,0}$ are the form factor slopes, I is the integral of 
the phase space and $\delta_{SU2}, \delta_{k}$ are the isospin-breaking
and electromagnetic radiative corrections; these corrections are
of the order of $\sim 1\div$2\%.
By measuring the BR($K_{l3}$) 
in a photon inclusive way  and correcting for the lifetimes 
the product $|V_{us}| f_{+}^{K^0\pi^-}(0)$ can be derived.
The four evaluations of this quantity from published data are in 
good agreement as shown in Fig.~\ref{fig:semilep}.b. On the other
hands, the recent measurement of BNL-E865~\cite{E865} gives a 
discrepant value which is instead consistent with unitarity and the
current determination of $|V_{ud}|$. Our preliminary measurement
of the BR$(\ks \ra \pi e \nu)$  allows us to obtain a new value of
$|V_{us}| f_{+}^{K^0\pi^-}(0)$ in much better agreement 
with E865 and unitarity (see Fig.~\ref{fig:semilep}.b).
The discrepancy between the $K_S$ and $K_L$, $K^{\pm}$ 
determination of $V_{us}$ calls for new measurements.
In the longer term, KLOE should be able to determine
all four $K_{l3}$ BR's to much better than 1\% and significantly 
improve the determinations of the lifetimes as well as the 
form factors slopes.

\section{Hadronic physics}
\begin{figure}[!t]
\begin{center}
\begin{tabular}{cc}
\epsfig{file=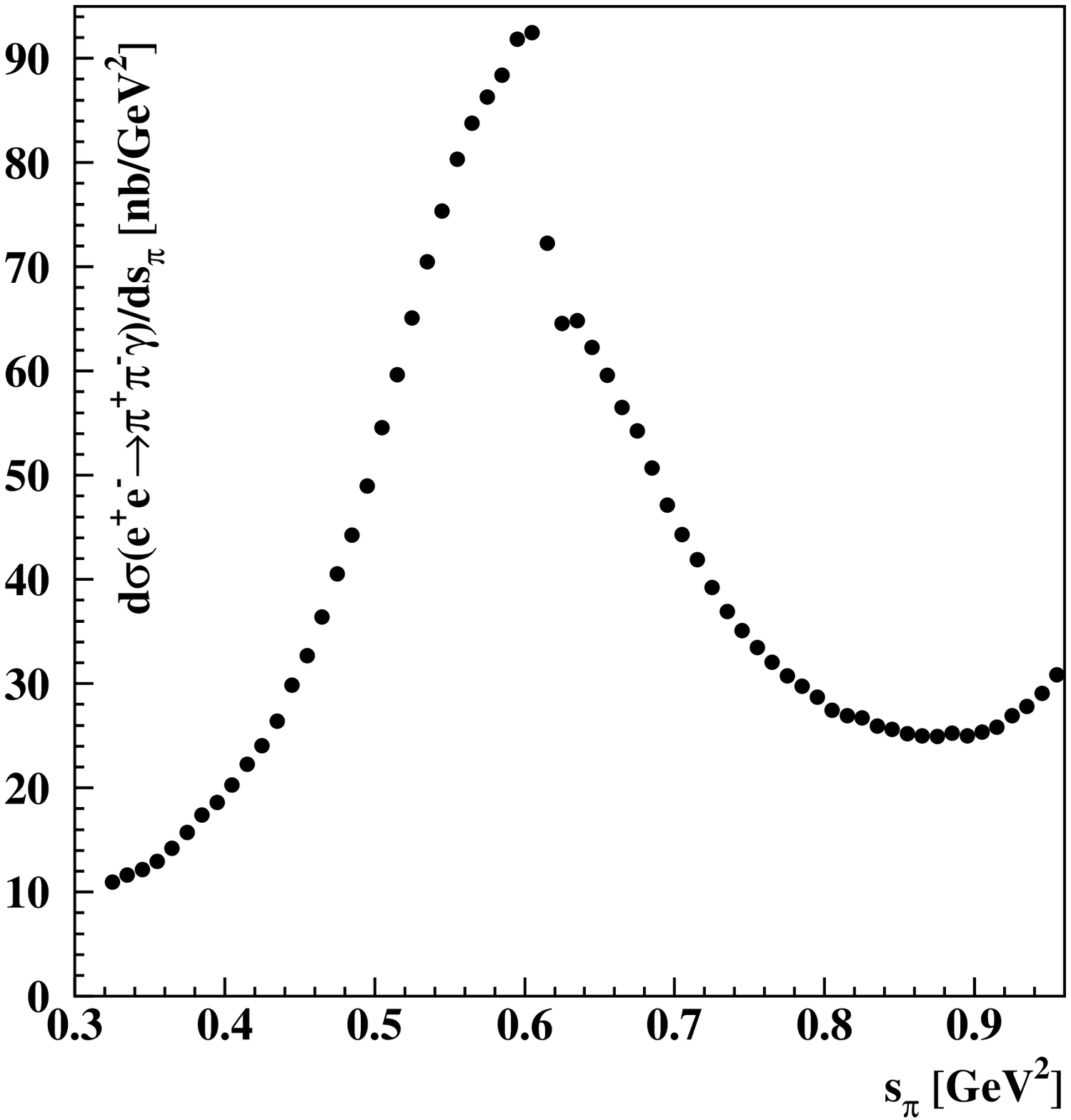,width=8cm}
\epsfig{file=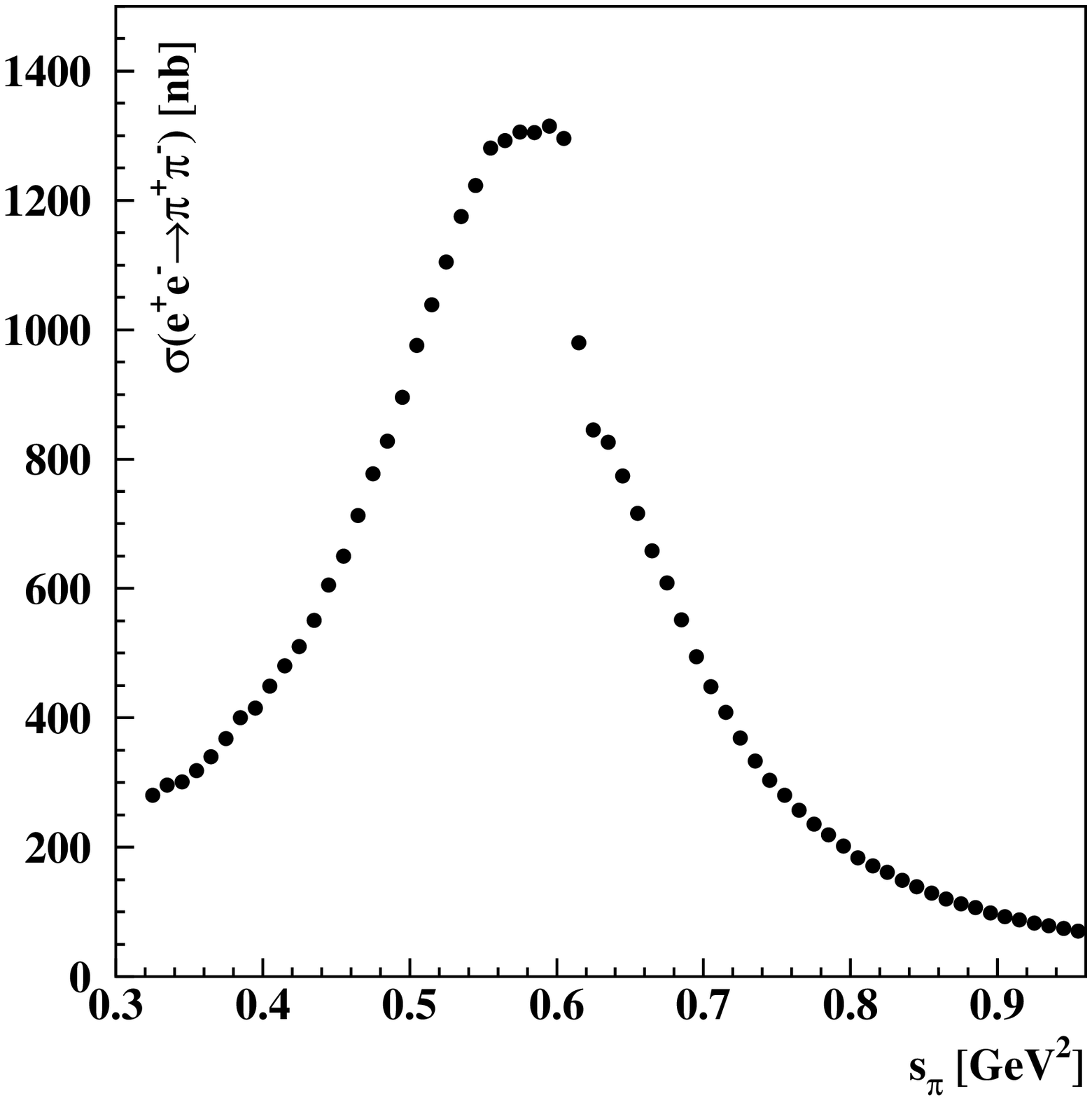,width=8cm}
\end{tabular}
\caption{ (Left) preliminary KLOE measurement of 
$d\sigma(e^+e^- \rightarrow \pi^+\pi^-\gamma)/ds_{\pi}$
for 140 pb$^{-1}$ of 2001 data; the cross section is 
inclusive in $\theta_\pi$ and with 
$\sin{\theta_{\gamma}}<\sin{15^o}$;
(Right) ``bare'' cross section for $e^+e^-\rightarrow \pi^+\pi^-$.
}
\label{fig:sigppg}
\end{center}
\end{figure}
Other than producing kaons, the $\phi$ meson decays $\sim$ 15\% of the
time in $\rho\pi$ and through radiative decays is a good source of
pseudoscalar ($\eta,\eta'$) and scalar ($f_{0}, a_{0}$) mesons.
Although a lot of interesting analyses have been published~\cite{eta1,f0,a0,rhopi}
on these items, and  their findings are  being improved with the 
larger statistical sample available, we do not discuss them here.

	The recent updated measurement of the anomalous
magnetic moment of the muon, $a_{\mu}$, by the E821 
collaboration~\cite{E821} has instead led to renewed interest in accurate
measurement of the hadronic cross section.
%
%
From the theoretical
side, the hadronic contributions on $a_{\mu}$, $a_{\mu}^{had}$, cannot
be evaluated in perturbative QCD but via a dispersion relation which
integrates the hadronic cross section multiplied by an appropriate kernel.
The process $e^{+}e^{-} \rightarrow \pi^{+} \pi^{-}$ below 1 GeV
accounts for $\sim$ 70\% to the $a_{\mu}^{had}$ value and of its error. 
The most recent measurement of 
$\sigma(e^+ e^- \rightarrow \pi^+ \pi^-)$  
by CMD-2~\cite{CMD2}, done with energy-scan  of $e^{+} e^{-}$ collisions,
claim statistical (systematic) precision of 0.7\% (0.6\%) and imply a
difference of $ - 2.7\,\sigma$ of the calculated value of
$a_{\mu}$ with respect to the  E821 measurement. Moreover, it gives
also a rather strong disagreement with the value of $a_{\mu}^{had}$
obtained using $\tau$-data~\cite{Davier} after isospin correction.

KLOE is determining in an original way this cross section as a function 
of $s_{\pi}$, the squared center of mass energy of the $\pi\pi$ system, 
in the region $0.3 < s_{\pi}<$ 1 GeV$^2$. DA$\Phi$NE operates at 
fixed energy $W \sim m_{\phi}$, but Initial State Radiation (ISR)
lowers the available beam energy for the di-pion system. 
We measure the cross section for the process 
$e^+e^- \ra \pip \pim \gamma$ and use the 
PHOKHARA generator~\cite{phokara} to relate $\sigma(\pip\pim\gamma)$ 
with $\sigma(\pip\pim)$.
Complications from processes with final-state radiation are
avoided by restricting the selection to events with small-angle photons
($\theta_{\gamma}<15^o$) where ISR events completely dominate the
sample. The $\gamma$'s are not detected, but $s_{\pi}$ and $\theta_{\gamma}$
are instead reconstructed by using DCH information on the $\pi$'s. 
A description of the
analysis strategy used can be found elsewhere~\cite{sigmahadeps}.
The preliminary KLOE data shown in Figs.~\ref{fig:sigppg} 
provide an independent measurement (also from
the systematic point of view) of this cross-section 
from CMD-2 data. We calculate the  dispersion integral 
in the same region used by CMD-2 ($0.37<s_{\pi}<0.93$ GeV$^2$) 
to get: $a_{\mu}^{had}=(376.5 \pm 0.8 \pm 5.9)\cdot 10^{-10}$.
This results is in good agreement with the CMD-2 number:
$a_{\mu}^{had}=(378.6 \pm 2.8 \pm 2.3)\cdot 10^{-10}$ 
confirming the discrepancy of $e^+ e^-$ data with $\tau$-data and
with the measured value of $a_{\mu}$.

\end{document}